\input{aipcheck}

\documentclass[
    ,final            % use final for the camera ready runs
  ]
  {aipproc}

\layoutstyle{8x11double}

\usepackage{color}
\newboolean{color}

\setboolean{color}{false}
\setboolean{color}{true}

\def \approxgt{\,\raise2pt \hbox{$>$}\kern-8pt\lower2.pt\hbox{$\sim$}\,}
\def \approxlt{\,\raise2pt \hbox{$<$}\kern-8pt\lower2.pt\hbox{$\sim$}\,}
\def \thi{\thinspace}
\def \ngth{\negthinspace}
\def \ngth2{\negthinspace\negthinspace}
\def \ni{\noindent}

\def \Teff{{$T_{\rm {ef\!f}} $}}

\def\Lo{{$L_\odot $}}
\def\Mo{{$M_\odot $}}

\def \eg{{{\sl e.g.},\ }}
\def \etal{{\sl et al.\ }}

\def \ie{{{\sl i.e.},\ }}

\def \viz{{\sl viz.\ }}

\def\nquad{\negthinspace\negthinspace\negthinspace\negthinspace}

\definecolor{firebrick}{rgb}{0.6,0.2,0.2}

\ifthenelse{\boolean{color}}
{\def\fb{\color{firebrick}}}{\def\fb{\color{black}}}

\begin{document}

\title{The State of Cepheid Pulsation Theory}


\author{J. Robert Buchler}{
  address={Physics Department, University of Florida,
Gainesville, FL 32611}
}

\begin{abstract} We review the current state of Cepheid modeling and discuss
its dominant deficiency, namely the use of time dependent mixing length.
Notwithstanding, Cepheid modeling has achieved some excellent successes, and we
mention some of the most recent ones.  Discrepancies between observations and
modeling appear not so much in the gross properties of single mode Cepheids,
but rather when more subtle {\sl nonlinear} effects are important, such as in
double mode or even triple mode pulsations.  Finally we discuss what we
consider the most important challenges for the next decade.  These are, first, 
realistic multi dimensional modeling of convection in a pulsating environment,
and second, the nonlinear modeling of the nonradial pulsations that have been 
observed, and, relatedly, of the Blazhko like phenomenon that has recently been
observed in Cepheids.
  \end{abstract}

\maketitle

%%%%%%%%%%%%%%%%%%%%%%%%%%%%%%%%%%%%%%%%%%%%
%% MAINMATTER
%%%%%%%%%%%%%%%%%%%%%%%%%%%%%%%%%%%%%%%%%%%%

\section{Introduction}

When looking back to the pioneering days of Cepheid modeling (\eg
\cite{christy64,cox66}) we have to modestly admit that in terms of numerical
modeling methods we have only made some progress, except that we now have at
our disposal enormous computing power compared to that available in those days.

There has however been quite a bit of progress in terms of the quality of the
opacities that are now used (OPAL \cite{iglesias90}, OP \cite{seaton94}, and
Andersen-Ferguson \citep{alexander94}).  In addition, some treatment of
convection is necessary and is routinely included in the calculations in the
form of time-dependent mixing length.

In parallel to direct numerical simulations, the {\sl amplitude equation
formalism} has been developed \cite{buchlergoupil84,dziembowski84,
takeuti,spiegel,coullet} (for a review see \cite{buchler93}).  The
physical conditions that prevail in Cepheids and RR Lyrae, namely that the
growth rates of the dominant modes are small compared to their frequencies,
form the basis for the applicability of these techniques.  We note in passing
that, in contrast, amplitude equations do not apply in Pop. II Cepheids, RV Tau
and Mira variables.  In these stars the growth rates are comparable to the
frequencies and therefore the amplitudes can change substantially on the time
scale of a period, hence the frequent occurrence of irregularity in the
pulsations \cite{chaos,reconstruction1,reconstruction2}.
 
In many respects the application of amplitude equations complements brute force
modeling.  Amplitude equations, or normal forms as they are known in nonlinear
dynamics \cite{guckenheimer}, are very general and describe the underlying
mathematical structure of the temporal behavior, \ie the pulsations in our
case.  In addition, when combined with numerical simulations \cite{kbby}, they
provide not only an excellent description of the modal selection problem, \eg
of where the region of double mode behavior occurs, or where hysteresis occurs
('either or' regime in the jargon of stellar pulsations), but they also yield a
deeper understanding of these bifurcations in the pulsational behavior
\cite{szaboetal}.

\section{Critique of the hydrocodes}

A number of hydrocodes have been developed over the last dozen years or so (\eg
the Italian code \cite{stellbono}, the Viennese code \cite{dorfifeuchtinger},
the Florida-Budapest code \cite{byk97} code.  They are all essentially the same
in that they include some similar form of time-dependent mixing length
approximation.  The first code is based on the formulation of \cite{stell}, and
the second on that of \cite{kuhfuss,gehmeyr}, and the third a hybrid of the two
formulations.  Some of their features have been compared in
\cite{buchlerkollathtc}.
More recent codes are the Australian \cite{woodolivier} and the
Polish ones \cite{smolec}.  Some of these codes include a moving adaptive mesh
that resolves better the sharp temperature features, in particular those
associated with the hydrogen partial ionization front
\cite{dorfifeuchtinger,marom}.  Some in addition improve on the equilibrium
heat diffusion \cite{dorfi}.

We briefly recall the ingredients of time dependent mixing length.  The
effects of turbulence and convection in 1-dimensional hydrodynamics description
appear through the turbulent pressure and viscous stresses $\fb p_t$ and $\fb
p_\nu$, the convective {\fb $F_c$} and the energy coupling term $\fb \cal{C}$.

\begin{eqnarray}
\nquad {du\over dt} \nquad&=&\nquad -{1\over\rho}{\partial \over\partial r}
\left(p+\thi\thi {\fb p_t}+{\fb p_\nu}\thi\thi \right)
   - {G M_r\over r^2}  \\
\nquad {d\thi e\over dt} +p\thi  {d\thi v\over dt}
 \nquad&=&\nquad -{1\over\rho r^2} {\partial \over\partial r} \left[ r^2
\left(F_r+\thi\thi {\fb F_c}\thi\thi \right)\right]
 +\thi {\mathbf{{\fb \cal C}}}
 \end{eqnarray}

The turbulent and convective quantities are assumed to be functions of the
'turbulent energy' ${\fb e_t}$ that is assumed to satisfy

\begin{equation}
   {d{\fb e_t}\over dt} +
    \left({\fb p_t}+{\fb p_\nu}\right)\thi  {d\thi v\over dt}
  = -{1\over\rho r^2} {\partial \over\partial r}\left( r^2 {\fb F_t}\right)
   -{\fb \cal C}
\end{equation}

The expressions for {\fb $p_t, F_t, p_\nu$} and 
{\fb $\cal{C}$} can be derived in
analogy with gas kinetic theory, however without the same solid physical basis,
or more simply just on dimensional grounds plus a few assumptions.
All these terms contain ($\alpha$) parameters of O(1) for whose values 
physics provides  little guidance.
There is also some ambiguity, \eg
possible physically acceptable choices for ${\fb F_c}$
and the source term
${\fb S_t}$ (that goes into the energy coupling term {$\fb\cal C$:})
\begin{eqnarray}
{\fb F_c} \sim {\alpha_c} \thi e_t^{{\scriptscriptstyle 1/2}}\thi  Y  &\quad&
{\fb F_c} \sim {\alpha_c}\thi  e_t \thi Y^{{\scriptscriptstyle 1/2}}
    \\
  {\fb S_t} \sim { \alpha_s}\thi  e_t^{{\scriptscriptstyle 1/2}}\thi  Y
  &\quad&
  {\fb S_t} \sim { \alpha_s}\thi  e_t\thi  Y^{{\scriptscriptstyle 1/2}}
\end{eqnarray}
\ni where
\begin{eqnarray}
Y \equiv \Big[-{H_p\over c_p} \thi\thi {\partial s\over \partial r}\Bigr]_+
\quad&{\rm or}&\quad
Y \equiv -{H_p\over c_p} \thi\thi {\partial s\over \partial r}
\end{eqnarray}

\ni More $\alpha$ parameters appear if a P\'eclet correction for radiative
losses
and a flux limiter for
 $F_c< \rho \thi c_p \thi T\thi  c_{sound}$ are included.

\begin{figure}
  \includegraphics[height=.4\textheight,width=.5\textheight]{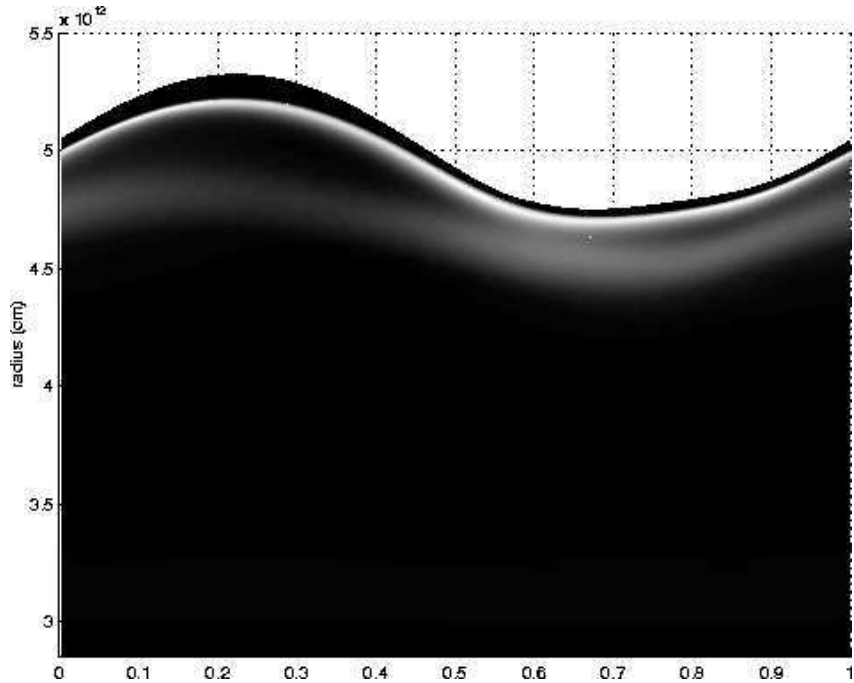}
\caption{Evolution of the specific 
   turbulent energy $e_t$ during the pulsational phase;
   the lightness of the grey reflects the strength of $e_t$}
\label{fig:et}
\end{figure}

There are a number of well known problems with time dependent mixing length.
To start, it is an empirical rather than a consistent physical description.
Second, the gradient approximations, \ie $F_t\propto\nabla e_t$ and
$F_c\propto\nabla s$, are poor \cite{canuto}.  In the same vein, 3D numerical
simulations show that the upward and downward plumes are highly non local
\cite{cattaneo91}, whereas mixing length assumes a local approximation
(diffusion of turbulent energy).  Third, some 8 or more {($\alpha$)}\thi
parameters appear in time dependent mixing length and physics provides no
guidance for their values.  Therefore some of these parameters are calibrated
with the help of a comparison of the results with observations, but the
parameter space is large!  The less important parameters are arbitrarily fixed.
It is somewhat disturbing that for best agreement with observations one needs a
different set of $\alpha$ values for RR Lyrae and for Cepheids, and also for
different metallicity.  Finally, there is no guarantee that time dependent
mixing length is flexible enough to describe convection sufficiently well,
despite the many {$\alpha$} parameters.

But these problems notwithstanding dependent mixing length has achieved many
successes.

Before going on to discuss these successes we want to display in
Fig.~\ref{fig:et} the behavior of the turbulent energy $e_t$ over a period in a
pulsating Cepheid model with a period of 10.9~days, and $M$=6.1\Mo,
$L$=3377\Lo, \Teff=5207K, $X$=0.70, $Z$=0.02 \cite{bykg}.  In the figure $e_t$
is shown as a function of Lagrangean radius and the lightness of the grey
reflects the strength of $e_t$.  The turbulent energy is largest in the region
associated with the combined H and first He ionization fronts which is the
convectively most unstable region, but it also appears in the next most
important, \viz the He$^{\scriptscriptstyle +}$--He$^{\scriptscriptstyle ++}$
region.  Fig.~\ref{fig:et} clearly demonstrates how the turbulent energy tracks
the source regions which move through the fluid during the pulsation.  It also
shows the importance of time dependence in the convective pulsating envelope.
The turbulent energy increases during the pulsational
compression phase and that, in this relatively hot model, the two turbulent
zones briefly merge.

\section{Recent results}

The literature is very vast, so we will just mention some of the more recent
work.  

One of the most striking properties of the Cepheids, both for fundamental (F)
mode pulsations and for first overtone (O1) mode pulsations is the progression
of the Fourier decomposition coefficients, of the light curves as well as of the
radial velocity curves (\eg \cite{soszynskifourier}).  For F mode Cepheids this
progression is of course related to the well-known Hertzsprung progression in
the bump Cepheids.  Full amplitude Cepheid model sequences do a good job at
reproducing the Fourier properties of the F Cepheids \cite{bono02} and of the
O1 Cepheids \cite{feuchtinger00}.

In parallel, the amplitude equation formalism that was developed for explaining
the effect of internal resonances on the appearance of the light and radial
velocity curves has given a clear demonstration that it is the 2:1 resonance
between the self-excited F mode and the vibrationally stable, but resonant
second overtone \cite{klapp,buchler90a} that is responsible for the structure
of the Fourier coefficients for periods around 10 days, rather than a shockwave
that reflects off the core.  For the first overtone Cepheids (or s Cepheids) it
is the 2:1 resonance of the stable fourth overtone with the self-excited first
overtone that causes structure at period in the vicinity of 4 days.
Modeling also reproduces well the observed light and radial
velocity curves of individual stars, \eg \cite{natale},
\cite{baranowski}.

A full amplitude model survey of F and O1 Cepheids \cite{szabo07} has found
that the phase lag between light and radial velocity curves is in good
agreement with observations.  This study has furthermore demonstrated that the
phase lag can also be used observationally as a discriminant between F and O1
mode pulsation.

The period ratio $P_1/P_0$ as a function of period $P_0$ of the beat Cepheids
is a sensitive function of the metallicity $Z$.  This property has recently
been taken advantage of to determine the metallicities in the LMC and
SMC \cite{marquette} with the help of Cepheid modeling
\cite{buchlermetall1,buchlermetall2}.  The same method has
been applied to the 5 known beat Cepheids in M33.  Interestingly, this yields a
galactic metallicity gradient for M33 that is in good agreement with totally
independent methods \cite{beaulieu}.
 
A recent study shows that a comparison of the light curves of full amplitude
bump Cepheid models with observed light curves can also be used as metallicity
tracers and distance indicators \cite{keller}.

Theoretical period-color-luminosity relations are in good agreement with
observational ones \cite{kanbur}, \cite{fiorentino}.  A summary of the recent
Italian work on this and other topics in Cepheid modeling appears in
\cite{caputograpes}.

Theory has been ahead of observations by predicting the existence of 'strange'
Cepheids and RR~Lyrae \cite{strange1,strange2}.  These are Cepheids or RR~Lyrae
that pulsate in a high (7th to 12th) overtone in which the pulsation is
confined to the outer region, more specifically, above the hydrogen ionization
front.  Typically, the predicted periods of these self excited modes are 4 to 5
times smaller than the fundamental period of the same object.  The amplitudes
are predicted to be in the millimag range.  On the observational side some
evidence for the existence of strange Cepheids has been uncovered.  However, it
is hard to distinguish between intrinsic pulsation and ellipsoidal binary motion at
the millimag level {\cite{ula1,ula2}).  Furthermore, in crowded field
conditions, it is possible that the objects could be regular, albeit low
amplitude Cepheids that appears above the P-L relation because of contamination
by another star.  Follow-up observations would be useful to ascertain that the
identified objects are indeed strange Cepheids.

In Fig.~\ref{fig:ula} we display the results of a recent search for ultra-low
amplitude (ULA) variability in the combined MACHO and OGLE LMC data
\cite{ula2}.  For reference and to guide the eye, the well known Cepheid P-L
strips are indicated as black dots.  All the ULA (Fourier amplitude $A <$ 0.01)
objects that were found in the plotted period--Wesenheit magnitude ($W = I -
1.55 (V-I)$) are shown as orange filled circles.  It is possible that some of
the objects are ellipsoidal binaries that with the available data, are
indistinguishable from ULA Cepheids.  The clustering of the ULA objects with
the LMC Cepheids suggests though that the latter are indeed pulsating
variables. The objects below the fundamental P--L relation are probably Pop. II
Cepheids.  The single object way above the P-L relation is a strange Cepheid
candidate.  When one concentrates on the remaining objects, their position is
seen to correlate strongly with the LMC Cepheid P-L relation.  One expects any
ULA Cepheids to lie at the edges of the instability strip, but
Fig.~\ref{fig:ula} might suggest that they lie on a parallel P-L relation that
is shifted slightly upwards.  If that is indeed the case, it is an interesting
challenge to explain the nature of these ULA Cepheids.

\begin{figure}
\ifthenelse{\boolean{color}}
  {\includegraphics[height=.5\textheight]{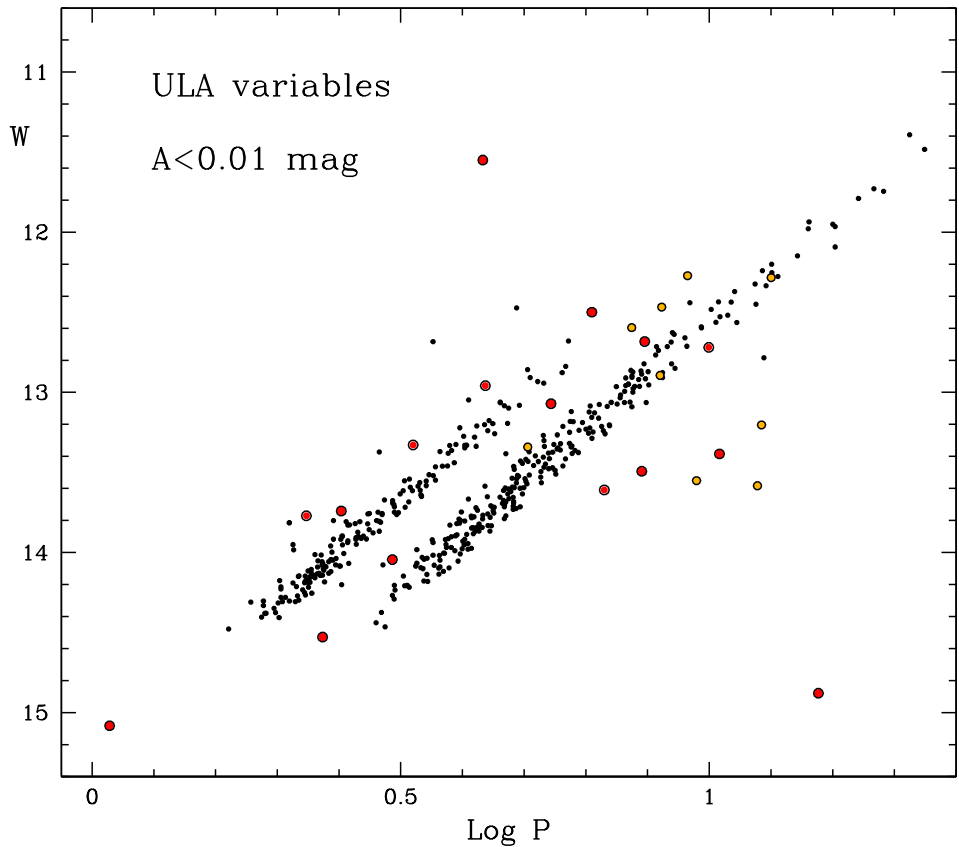}}
  {\includegraphics[height=.5\textheight]{plot_ula_cephs.ps}}
  \caption{Ultralow amplitude objects in the LMC, shown as open
      (red and orange) circles.
  The smaller (orange) circles possibly have proper motions.  The classical
  Cepheids are superposed as dots for reference.}
\label{fig:ula}
\end{figure}

The analyses of the OGLE data base {\eg
\cite{soszynskifourier,soszynski08,moskalikkola}), to which we refer the reader
for the following discussion, have yielded a beautiful overview of the HR
diagram.  In addition to the F and O1 Cepheids, a whole zoo of other types of
Cepheid pulsational behavior has now been observed, namely O2 single mode,
double mode F/O1, double mode O1/O2, triple mode, and Cepheids with Blazhko
like modulations.

\begin{figure}
\ifthenelse{\boolean{color}}
  {\includegraphics[height=.56\textheight]{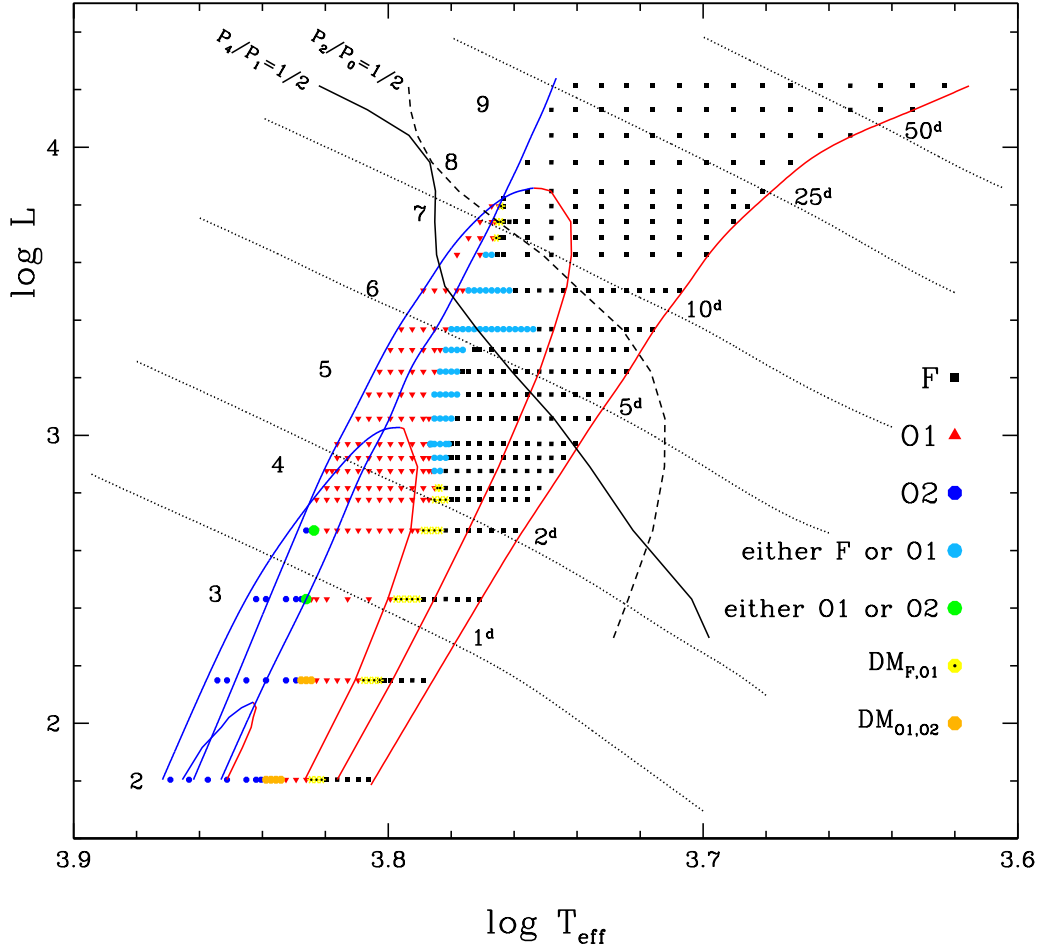}}
  {\includegraphics[height=.56\textheight]{HR_gal_full_bw.ps}}
\caption{Computed HR diagram indicating the domains of single mode pulsations,
  of double mode pulsations and of hysteresis}
\label{fig:HR}
\end{figure}

For comparison we present the results of a recently computed HR diagram
\cite{szabobuchler} in Fig.~\ref{fig:HR}.  The bounding curves on the right [in
red, in the color figure of the electronic version] and the [blue] ones on the
left represent respectively the red and blue edges of the F, O1, O2 and O3
pulsational instability strips.  We recall in passing that these boundaries are
obtained from linear vibrational stability analysis, and as such, they only
tell us where at least one mode is linearly unstable, \ie where pulsations
occur, but they generally do not tell us what type of pulsation actually
results.  For example, for F/O1 double mode behavior it is {\sl necessary,
but not sufficient} that both modes be linearly unstable (in the figure the
intersection of the F and the O1 instability strips).  Nonlinear calculations
are necessary to determine which pulsational behavior can actually occur.  (We
say 'can occur' rather than 'occurs' because there may also be hysteresis, \ie
the pulsational state depends on the evolutionary path.)

Each of the symbols in Fig.~\ref{fig:HR} represents the result of one or more
hydrodynamic integrations to steady full amplitude.  The numbers on the left
denote the mass (in \Mo) which is constant along horizontal lines.  The
fundamental periods (in days) are constant along the dotted lines.
The open squares [black] denote the regime where only single mode F full
amplitude pulsations (limit cycles) occur, the open upside-down triangles [red
filled triangles] the regime of single mode O1, and the open circles [blue
filled circles] that of single mode O2 pulsations.

The solid [lightblue] squares denote the regime where either F or O1 pulsations
occur, \ie where there is hysteresis.  A star, say with mass 5\Mo, evolving
leftward enters the F instability strip (rightmost red line) and starts to
pulsate in the F mode.  It continues to do so to the leftmost edge of the
(lightblue) hysteretic region where it switches to O1 pulsations until it exits
through the O1 blue edge on the left.  On its return, moving now rightward, it
enters the O1 instability strip (leftmost blue line) and starts to pulsate in
the O1 mode.  It continues to do so to the rightmost edge of the (lightblue)
hysteretic region where it switches to F pulsations until it exits through the
F red edge on the right.

Below, and as a continuation of the hysteretic region we typically find F/O1
double mode pulsations denoted by crosses [yellow filled circles with a
concentric black dot].  Care has to be exercised in the modeling computations
that the pulsations are truly double mode (\ie with constant amplitudes and
phases) rather than that the model may be switching from one mode to the other.
Indeed, it has been shown that the amplitudes may level off relatively fast,
and then vary extremely slowly, giving the erroneous impression of steady
pulsations.  To be certain that double mode is indeed obtained it is necessary
to integrate the model with several judiciously chosen initial kicks and to
watch the evolution of the individual amplitudes in an amplitude--amplitude
portrait \cite{szaboetal}.

Interestingly, F/O1 double mode pulsations can also occur in a narrow region
above the hysteresis regime.  This double mode behavior is induced by the $P_0
= 2 P_2$ resonance in the vicinity of the 10\thi d period.  The locus of the
center of this resonance is indicated as a dashed line in the figure.

The width of the hysteresis regime [light blue lines] is clearly nonmonotone
with mass.  This is not a numerical problem, but we have again traced this
behavior to a resonance, this time $P_1 = 2 P_4$, as seen by the locus of the
resonance center (solid line).

Similar regimes of hysteresis and double mode occur on the right towards the
lower masses, but now for O1 and O2.  The hysteretic regime of O1 and O2
('either O1 or O2') is marked by solid triangles [green filled circles].  Below
this occurs a regime of O1/O2 double mode pulsations labeled with x's
[orange filled circles].

We note in passing, and as a curiosity, that we have also found an extremely
narrow strip of hysteresis with either F single mode pulsations or F/O1 double
mode pulsations.

It is likely that multimode and hysteresis also occur for
the 2\thi\Mo\ models, for which O3 is linearly unstable, but we have
not checked this.

Overall, the modeling is seen to reproduce the different types of single mode
and double mode behavior with approximately the right topography.  However,
some types of observed pulsational behavior, such as triple mode, have yet to
be modeled.  It is clear that double mode or triple mode behavior is much more
sensitive to the details of the physical input than, say single mode, because
it depends on more subtle effects, \viz the nonlinear coupling between the
modes.

\vskip 5pt

Before going on we also want to mention some well known, but serious
discrepancies, for which stellar evolution calculations are at fault, but on
which Cepheid modeling unfortunately has to rely.

First, low mass, low Z evolution loops do not penetrate the instability strip
where Cepheids are actually observed.  It has been suggested that this could be
a metallicity selection effect \cite{cordier,kovacs09}.  There remains
disagreement about the treatment of convection and convective overshoots which
leads to uncertainties in the Cepheid M-L-Z relation.  Unfortunately, this in
turn causes uncertainties in Cepheid modeling which depends on an ML relation.
However, we note that the modeling of LMC and SMC Cepheids
\cite{buchlerkollathbeaulieu} based on the tracks of Girardi \etal
\cite{girardi} is in good agreement with the resonance constraints imposed by
observations.  Finally, the occurrence of a bump mass discrepancy was largely
removed with the OPAL opacities \cite{moskalikbuchler}, some authors claim that
there remains a small discrepancy \cite{keller} which is at variance with the
above mentioned work \cite{buchlerkollathbeaulieu}.

\section{Beyond time-dependent mixing length}

Since dependent mixing length has such a long list of known problems, efforts
have been made to go beyond the simple model described above, while keeping its
1D feature.  This unfortunately leads to a proliferation of equations
\cite{canuto,gough,kupka,houdek} with a concomitant numerical cost and perhaps
little guarantee  of substantial improvement.

In parallel, plume models have been proposed \cite{rieutord}.  Along similar
lines Stoekl has developed a 2 column model \cite{stoekl}.  However, none of
these approaches have been used in Cepheid models, nor are they expected to
provide a parameter free description of convection which after all is 3
dimensional phenomenon.

\section{Two Grand Numerical Challenges}

In our opinion there remain two grand challenges for the modeling of Cepheids,
and for that matter of RR Lyrae which are quite similar in many ways.  These
are the modeling of convection in the highly structured envelopes of Cepheids,
on the one hand, and of nonradial pulsations, on the other.

\subsection{1. Multidimensional 
simulations of convection in Cepheids and RR Lyrae}

There is a great deal of literature on the numerical difficulties that arise in
the modeling of astrophysical convection.  The literally astronomically large
Rayleigh numbers imply that the turbulence is well developed.  This in turn
implies that many degrees of freedom are involved which sets extreme mesh
requirements.  However, the hope of all large eddy simulations (LES) is that
once we have attained a sufficient spatial resolution, then the dominant, large
scale features of convective transport are well reproduced.  Another problem
arises from the extremely small Prandtl number.  (The Pr number is the ratio of
gas viscosity to heat conduction.)  In the numerical modeling the viscosity is
however not set by the physical extremely small viscosity, but by that of the
numerical scheme.  Effective Pr numbers of order 0.01 or less are hard to
achieve.

There has been a great deal of excellent multi-dimensional numerical
hydrodynamics work in other astrophysical contexts, such as solar models,
stellar core burning, dynamos \cite{stein,arnett}. Additional difficulties
arise that are specific to Cepheids and RR Lyrae modeling.  First, it is out of
question to compute convection over the whole sphere, nor is it probably
necessary.  If convection occurs over a vertical range $\Delta R_c$ it seems
reasonable that we limit ourselves to a horizontal sector whose width is a
few times $\Delta R_c$ at the vertical center $R_c$ of the convective region,
\ie an angular sector $\Delta\theta= \Delta R_c/R_c$, typically a degree or so,
and impose periodic horizontal boundary conditions.  But even this
leaves extreme mesh requirements.  The resolution of the rapidly varying
temperature structure in the partial hydrogen ionization region requires tiny
$\Delta R$.  However, the hot inner region that needs to be included in a
realistic stellar model, even though no convection is taking place there
requires large $\Delta R$ because the sound speed is large, so that we are not
killed by the Courant condition.  The mesh aspect ratios can thus vary from
0.01 to 100.

Second, the surface boundary condition is delicate because the envelope is
surrounded by an optically thin, approximately isothermal atmosphere with an
exponential decrease of density.

Third, the pulsating environment in which the stellar radius can change by more
than 10\%, combined with the exponentially decreasing atmosphere, make an
Eulerian code unsuitable.  A moving mesh is needed to follow the average
pulsational motion, and adaptive features are required to track the sharp
temperature gradients (H ionization front).

It should be clear that the modeling of convection in a pulsating star is a
challenging problem that is at the limit of current modeling possibilities.
However an intermediate goal could be to make a small but sufficient number of
3D simulations in realistic Cepheid or RR Lyrae models, and then use the
results to calibrate the mixing length recipe parameters, or if necessary to
improve or even replace mixing length theory.

\subsection{2. The Modeling of Nonradial Pulsations in Cepheids}

The analysis of the observations of the OGLE Cepheids strongly suggest the
excitation of nonradial modes \cite{soszynski08,moskalikkola}.

On the theoretical side, the linear stability analysis of nonradial modes in
Cepheids and RR Lyrae is notoriously difficult \cite{dziembowski},
\cite{osaki}.  The problem arises from the fact that the nonradial modes have
mixed p and g mode nature with unresolvably rapid spatial oscillations in
interior.  This means that most nonradial modes are very strongly damped.
However, some surface-trapped modes were found to be unstable or only mildly
damped, and they can therefore compete dynamically with the low order radial
modes \cite{balona,vanhoolst}.  A recent revisit of the problem with realistic
stellar models \cite{mulet} however finds no nonradial instability among the
low $\ell$ modes. (Only low $\ell$ are detectable with full disk observations.)
This problem therefore remains a puzzle even at the linear level.  Ultimately,
of course, nonlinear 3D hydrodynamic simulations will be necessary to model the
observed nonradial pulsations, which is again a very tough numerical problem,
for many of the same reasons as for convection

The analyses of the LMC variables \cite{soszynski08,moskalikkola} also show
the existence of Blazhko like modulations of the pulsations which bear some
similarities with the same effect in RR Lyrae.  Just as in the latter the
physical origins remain obscure.  Similar modulations also seem to occur in 
in V473 Lyr and in Polaris \cite{bruntt}.

Among the viable models for the Blazhko effect are the likely involvement of
one or more nonradial modes \cite{coxnr}.  A specific mechanism involving 
resonant coupling has been examined by \cite{kovacsreson}.

One may also wonder whether the nonradial modes can be destabilized by
convection?  Could this be the origin of the observed Blazhko modulations
\cite{moskalikkola}?  We are somewhat skeptical that a simple dependent mixing
length description is sufficiently good to answer that question.  Most likely a
full 3D simulation, as described in the previous section, will be necessary to
treat the convection -- pulsation interaction.  There is also the possibility
the Blazhko modulations arise from the radial pulsation - convection
interaction.  In any case an explanation of the physical mechanism of the
Blazhko cycle remains a theoretical challenge.

\vskip 10pt

Despite these serious difficulties, we think that these two grand modeling
challenges will be attacked and possibly will be met in the next decade.

\begin{theacknowledgments}
We gratefully acknowledge the support of the National Science Foundation
(grant AST 07-07972).  We also thank Robert Szab\'o for letting us show the
computed HR diagram prior to publication.
\end{theacknowledgments}

%%%%%%%%%%%%%%%%%%%%%%%%%%%%%%%%%%%%%%%%%%%%%%%%%%%%%%%%%%%%%%%%%%%%%%%%%%%

\end{document}